\documentclass[12pt]{iopart}
\usepackage{iopams}

\begin{document}

\title[]{Statistical mechanics of generally covariant quantum 
theories: A Boltzmann-like approach}
\author{Merced Montesinos\dag\footnote[4]{e-mail: merced@fis.cinvestav.mx} 
and Carlo Rovelli\P\ddag\footnote[9]{e-mail:rovelli@cpt.univ-mrs.fr}} 


\address{
\dag\  Departamento de F\'{\i}sica, Centro de Investigaci\'on y de
        Estudios Avanzados del I.P.N., Av. I.P.N. No. 2508, 07000
        Ciudad de M\'exico, M\'exico.}

\address{
\P\ Department of Physics and Astronomy, 
University of Pittsburgh, Pittsburgh, PA 15260, USA.}

\address{
\ddag\ Centre de Physique Theorique, CNRS, Luminy, F-13288 
Marseille, France.}

\begin{abstract} 
We study the possibility of applying statistical mechanics to
generally covariant quantum theories with a vanishing Hamiltonian.  We
show that (under certain appropriate conditions) this makes sense, in
spite of the absence of a notion of energy and external time.  We
consider a composite system formed by a large number of identical
components, and apply Boltzmann's ideas and the fundamental postulates
of ordinary statistical physics.  The thermodynamical parameters are
determined by the properties of the thermalizing interaction.  We
apply these ideas to a simple example, in which the component system
has one physical degree of freedom and mimics the constraint algebra
of general relativity.
\end{abstract} 

\maketitle

\section{Introduction}

General relativity has modified our understanding of the
physical world in depth and has altered some among the most
fundamental notions we use to describe it.  During the last
ten years, the effort to understand the combined
consequences of this conceptual revolution and quantum
mechanics has lead to loop quantum gravity, a predictive
quantum theory of the gravitational field, whose theoretical
results can be, in principle, empirically tested
\cite{Rovelli:1998}.  There are other areas in our
understanding of nature, however, where the consequences of
the general relativistic conceptual revolution have not been
fully explored yet.  Among these is statistical mechanics. 
To be precise, thermodynamics and statistical mechanics on a
fixed curved spacetime have been much studied (see, for 
instance, \cite{tolman}); but not much is known on the
possibility of developing thermodynamics and statistical
mechanics of a fully general covariant system, in
particular, a system including the gravitational field.

Here, we begin to address this issue.  Specifically, we
study the following problem.  Consider a simple physical
system, $s$, with a finite number $D$ of degrees of freedom. 
Assume that $s$ is described by a fully constrained
Hamiltonian system.  That is, its dynamics is not given by a
Hamiltonian, but rather by $M$ first class constraints. 
Physically, this means that we do not understand the
dynamics of $s$ in terms of the evolution of $D$ dependent
Lagrangian variables (or $2D$ phase space variables) as
functions of a single preferred independent external time
variable $t$; rather, we understand the dynamics as the
relative evolution of $2(D+M)$ phase space variables with
respect to one another -- all the variables being on the
same footing.  The dynamics fixes relations between these
variables, so that by knowing some of them we can predict
the others.  Such a simple system encodes an critical
feature of general relativistic systems: the absence of a
preferred time variable, and the relational aspect of
evolution.  Now, consider a macroscopic system $S$ composed
by a large number of component systems, each one identical
to $s$, and interacting weakly.  Can we use statistical
mechanics to describe macroscopic properties of $S$?  Notice
that there is no time variable in the description of $S$,
therefore no notion of thermalization `in time'; there is
also no notion of energy, and thus no obvious way to define
a canonical or microcanonical ensemble.  If we arbitrarily
choose one variable in $S$ as the physical time (that is, if
we `deparameterize' the system), and then use conventional
statistical techniques, our results are going to depend 
on the choice of time, and therefore to be possibly
unphysical.  Is there anything we can nevertheless say,
about the macroscopic behavior of this system?  Can we still
apply thermodynamical or statistical mechanical techniques?

These questions are relevant in a strong-field gravitational
context, whenever a preferred time and a conserved energy are not
defined.  Of course, if we consider a system with a notion of
time and with conserved energy, we expect that temperature and
energy will recover their traditional role.  This is the case,
for instance, of an asymptotically flat gravitational field; in
this case the Hamiltonian is given by suitable boundary terms and
the observables at infinity evolve in the Lorentz time of the
asymptotic metric.  The general theory we present here will have
to yield standard results in this case. However, what about the
situations in which there is no conserved energy and no preferred
time?  For instance, as far as we know, our universe might very
well not be asymptotically flat.  Alternatively, we may be
interested in a system with a strong (dynamical) gravitational
field, and have no access to an external asymptotic region.  In
particular, consider a ``high temperature" early-universe regime. 
This is usually described in terms of fluctuations around a
background metric; is there a genuinely general covariant
description of this physics?  And what is temperature in this
context, if we do not fix a background metric?  Certainly, it is
difficult to even define what statistical mechanics is if we do
not have some notion of energy conservation; but does this mean
that in all gravitational systems in which there is no conserved
energy (most of them!), we have to renounce using statistical
methods?
   
These questions have not yet been addressed in the literature, as
far as we know.  An attempt to study certain aspects of the
foundations of general covariant statistical theory is in
Refs.~\cite{Rovelli:1993,Connes:1994}.  In these works, the
question addressed is whether a preferred time flow, having the
thermodynamical properties that we ascribe to physical time, can
be derived from the statistical mechanics of a covariant system. 
The answer is positive, and the flow turns out to be dependent on
the statistical state.  The relation flow/state reflects a very
general operator algebra structure (Tomita-Takesaki theorem), and
raises intriguing physical issues, in particular in view of
powerful mathematical uniqueness results about the flow (Connes'
Cocycle Radon-Nikodym theorem).  Here, on the other hand, we are
not concerned with the emergence of a time flow.  Instead, we
address directly the issue of a statistical description
independent from any notion of time.

Furthermore, Refs.~\cite{Rovelli:1993,Connes:1994} take Gibbs' 
\cite{Gibbs} (and Einstein's \cite{Einstein}!)  point of view on 
statistical mechanics: a statistical state is described by a 
distribution over the phase space $\Gamma$ of the compo\-site 
system $S$ (in Ehrenfest's terminology, over the $\Gamma$-space 
\cite{Ehr}).  The state represents the distribution of $S$'s 
microstates over many imaginary copies of the system, all in the 
same macrostate.  Here, on the contrary, we use Boltzmann's 
original point of view \cite{Boltzman}: we assume that $S$ is 
composed by a large number of identical subsystems $s$.  The 
statistical state is then described by a distribution over the 
phase space $\gamma$ of the component system $s$ (over the 
$\mu$-space, in Ehrenfest's terminology).  This gives, for each 
state of $s$, the expected number of component systems that are 
in that state.

Of course, we do not expect any of the well known subtleties and 
conceptual difficulties of statistical mechanics to be solved by 
applying it to covariant systems.  Here we are not concerned with 
the old problems in the foundations of statistical mechanics, but 
only with the specific new problems --and new beauties-- that 
emerge in trying to extend the general relativistic revolution to 
statistical physics.

Our main result is the following.  We argue that, under appropriate
conditions, the statistical mechanics of a system $S$ composed by many
constrained systems $s$ is well defined.  In particular, statistical
mechanics is not necessarily tied to the concept of energy, or to a
preferred time flow.  Accordingly, general covariant statistical
mechanics is not governed by the notion of temperature.  Instead,
intensive macroscopic parameters are determined by the properties of
Boltzmann's thermalizing interaction among the individual component
systems.  In the course of the paper, we develop the basis of
covariant quantum statistical mechanics and define the intensive and
extensive thermodynamical quantities.  

We begin by recalling the properties and the physical interpretation
of the parameterized systems in Section 2.  We then give the main
discussion on the foundations of covariant statistical mechanics in
Section 3, and a simple example in Section 4.  We discuss the
statistical mechanics of a gas of free relativistic particles in 
Section 5, and we comment and summarize in Section 6. 


\section{Presymplectic systems}

We consider fully constrained systems, with a finite number of 
degrees of freedom, and with first class constraints 
\cite{Dirac}.  Their dynamics is obtained from the action
\begin{eqnarray}
S [ q^i , p_i , \lambda^m ]& = & \int d \tau \,\, \left \{
\frac{dq^i}{d\tau}p_i - \lambda^m C_m (q^i,p_i) 
\right \} \, ,
\label{act}
\end{eqnarray} 
which is invariant under arbitrary reparametrizations of the 
parameter $\tau$.  The parameter $\tau$ is unphysical and 
unobservable, like the time coordinate in general relativity.  
The unreduced, or extended phase space $\gamma_{ex}$ is 
coordinatized by the canonical pairs $(q^i , p_i)$; $ 
i=1,2,...,N$.  The canonical 2-form on $\gamma_{ex}$ is 
$\omega_{ex}=dp_i \wedge d q^i$.  The pair $(\Gamma _{ex}, 
\omega_{ex})$ forms a symplectic space.  The variation of the 
action with respect to the canonical coordinates $q^i , p_i$ 
gives the equations of motion
\begin{eqnarray}
\frac{dq^i}{d\tau} & = &  \lambda^m 
\frac{\partial C_m (q^i ,p_i)}{\partial p_i} \, , \nonumber\\
\frac{dp_i}{d\tau} & = & - \lambda^m 
\frac{\partial C_m (q^i ,p_i)}{\partial q^i} \, ,
\end{eqnarray}
while the variation of the action with respect to the Lagrange 
multipliers $\lambda^m$ gives the constraint equations
\begin{eqnarray}
C_m & = & C_m (q^i ,p_i) = 0 \, , \quad m=1,2, ... , M  
\, .\label{firstclass}
\end{eqnarray}
Thus, the dynamics of the system with respect to $\tau$ is the 
unfolding of the gauge symmetry generated by the first class 
constraints, i.e., dynamics is gauge.

The first class constraints satisfy, in general, a ``non-Lie''  
algebra
\begin{eqnarray}
\{ C_m , C_n \} & = & C_{mn}\,^l (q^i , p_i)\ C_l \, ,
\end{eqnarray}
and the number of independent physical degrees of freedom of the 
theory is $D=N-M$.  The constraint surface $\gamma$ in 
$\gamma_{ex}$ defined by the constraint equations 
(\ref{firstclass}) is a $(2D + M)$-dimensional manifold.  The 
restriction $\omega$ of $\omega_{ex}$ to the constraint surface 
$\gamma$ is of rank $2D$.  The $M$ null directions of $\omega$ 
are the infinitesimal transformations generated by the 
constraints.  They define the gauge orbits on $\gamma$.  The 
physical phase space $\gamma_{ph}$ is the space of these orbits.  
This is the space of the physically distinct solutions of the 
equations of motion.

The space $(\gamma,\omega)$ is a presymplectic space, which 
contains the full dynamical information about the system.  Hence 
dynamical systems in this form are also called `presymplectic 
systems'.  $\gamma$ can be parameterized by the set of 
independent coordinates $(\tilde q^{a}, \tilde p_{a} , t^m)$, 
where $(\tilde q^{a} , \tilde p_{a}),\ a =1, 2, ..., D$ are 
canonical variables that coordinatize the physical phase space 
$\gamma_{ph}$, and $t^m ,\ m=1, 2, ..., M$ coordinatize the 
orbits.  In general this coordinatization can hold only locally, 
and different charts may be needed to cover the entire space.

Any conventional dynamical system with phase space $(\gamma_{ph}, 
\ \omega_{ph} = d\tilde p_a \wedge d\tilde q^a)$, and Hamiltonian 
$H=H(\tilde p_{a}, \tilde  q^{a})$ 
can be represented as a presymplectic system 
as
\begin{equation}
    (\gamma=\gamma_{ph}\times R,\ \  
    \omega=\omega_{ph}-H(\tilde p_{a}, \tilde  q^{a})\wedge dt),
\label{pres}
\end{equation}
where $t$ is the coordinate in $R$, and corresponds to the 
external time variable.  The difference between the conventional 
formulation and the presymplectic formulation is only in the fact 
that this time variable is treated on the same footing as the 
other variables.  As a concrete example, we may imagine that $H$ 
is the harmonic oscillator Hamiltonian describing the small 
oscillations of a pendulum, while $t$ is the reading of a 
physical clock.  Then the presymplectic system (\ref{pres}) 
describes how two equal-footing physical variables (the pendulum 
amplitude and the clock reading) evolve with respect to one 
another.  In general covariant systems, such as any general 
relativistic system, this `equal footing' status between all 
physical variables is an essential feature of the theory.  It 
expresses the major physical discovery of general relativity: the 
complete relativity of spacetime localization.

Note that the canonical coordinates $\tilde q^{a}$, and 
$\tilde p_{a}$ {\it are} the physical observables of the system.  
They are gauge-invariant.  They satisfy $\{ \tilde q^{a}, \tilde 
p_{b} \}=\delta_{b}^{a}$ on the physical phase space.  In these 
coordinates, the physical symplectic form on $\gamma_{ph}$ is 
$\omega_{ph}= d {\tilde p}_{a} \wedge d {\tilde q}^{a}$.  The general 
solution of the equations of motion is simply given by the embedding  
equations of the orbits in $\gamma_{ex}$, that is
\begin{eqnarray} 
q^i & = & q^i (t^m; \tilde q^{a}, \tilde p_{a}) \, , \label{ECI}\\
p_i & = & p_i (t^m; \tilde q^{a}, \tilde p_{a}) \, \label{ECII}.
\end{eqnarray}
Each set $(\tilde q^{a}, \tilde p_{a})$ determines a solution; 
along each solution, the quantities $(q^i, p_i)$ depend on the 
$M$ parameters $t^{m}$ (instead than just on a single time 
variable) because of the gauge freedom in the evolution. The inverse 
relations of (\ref{ECI})-(\ref{ECII}) give the 
dependence of the physical observables $\tilde q_{a}$, and 
$\tilde p_{a}$ from the original coordinates 
\begin{eqnarray}
\tilde q^{a} & = & \tilde q^{a} (q^i, p_i) \, \label{OI},\\
\tilde p_{a} & = & \tilde p_{a} (q^i, p_i) \, \label{OII},
\end{eqnarray}
as well as the orbit coordinates $t^m$
\begin{eqnarray}
t^m & = & t^m (q^i ,p_i)\, . \label{GI}
\end{eqnarray}
The quantities (\ref{OI}) and (\ref{OII},) commute with all 
the constraints, and provide a complete set (in the sense of 
Dirac) of gauge-invariant observables.  Every other physical 
observable can be obtained from them.  

Let us recall how evolution can be obtained from the basic 
observables (\ref{OI}) and (\ref{OII}) \cite{evolution,Mon2000}.  If 
we plug the gauge variables (\ref{GI}) into the full solution 
(\ref{ECI}) and (\ref{ECII}) we obtain the equations
\begin{eqnarray}
q^i & = & q^i (t^m (q^i, p_i); \tilde q^{a}, \tilde p_{a}) \,,\\
p_i & = & p_i (t^m (q^i, p_i); \tilde q^{a}, \tilde p_{a}) \, .
\end{eqnarray}
In general, $2N-M$ of these equations are independent.  For each 
physical state of the system, determined by the value of 
$({\tilde q}^a , {\tilde p}_a)$, these equations define an $M$ 
dimensional subspace in the phase space.  Therefore each state 
determines a set of relations on the original phase space 
variables.  These relations represent the dynamical information 
on the system; they provide the full solution of the dynamics in 
a {\it gauge-invariant} fashion \cite{Mon2000}.  

In particular, we might arbitrarily choose a set of M coordinates 
$q^m$ (or momenta $p^m$; or a combination of both) as independent 
`clock and position' variables, and express the evolution of the 
remaining set of coordinates and momenta as functions of these 
$q^{m}$ for any physical state $({\tilde q}^a , {\tilde p}_a)$.  
For each fixed numerical value $\hat q^{m}$ of the coordinates 
$q^m$, we have a well defined gauge-invariant observables in 
$\gamma_{ph}$.  For instance, let us chose (arbitrarily) $q^{1}$ 
as a dependent `partial' observable\footnote[3]{A `partial' 
observable is a physical quantity to which we associate a number, 
such as time $t$, position $x$ or electric field $E$.  A 
`complete observable' is a physical quantity that can be 
predicted if the state is known, or, equivalently, that gives us 
information on the state, for instance the value $E(t,x)$ of the 
electric field in a certain point $x$ at a certain time $t$.  For 
the notions of partial observable and complete observable, see 
\cite{boston}.}, and the next $M$ of the $q^{i}$'s, as 
independent `partial' observables, or `clock and position 
variables'.  That is, let us choose $m=2,\ldots ,M+1$.  Pick $M$ 
fixed numerical values $\hat q^{m}$ for the $M$ variables 
$q^{m}$.  Generically, this fixes uniquely a point on every 
orbit.  The value $Q^{1}_{\hat q^{m}}$ of $q^{1}$ on this point 
depends on the orbit, and can be obtained from 
(\ref{OI}-\ref{GI}).  Let it be
\begin{equation} 
Q^1_{\hat q^{m}} = Q^1_{\hat q^{m}}(\tilde q^{a}, \tilde p_{a}) 
\label{ec}.
\end{equation}
Here all the $q^{i}$ are partial observables, while $Q^1_{\hat 
q^{m}}$ is a complete observable.  The function (\ref{ec}) is 
gauge-invariant, well defined on $\gamma_{ph}$ and expresses the 
relative evolution of $q^{1}$, as a function of the $q^{m}, 
m=2,\ldots ,M$.  It is called an `evolving constant of the 
motion', or simply a `relational observable' \cite{evolution}.

The quantum theory can be constructed by imposing the quantum 
constraints on the unconstrained Hilbert space ${\cal H}$ (or 
some suitable extension of the same if the constraints have 
continuum spectrum).  The space of solutions of the constraint 
equations is the physical Hilbert space ${\cal H}_{phys}$ of the 
theory.  (If ${\cal H}_{phys}$ is not a subspace of ${\cal H}$ a 
scalar product is determined in ${\cal H}_{phys}$ by the 
requirement that the self-adjoint observables in ${\cal H}$ which 
are well defined on ${\cal H}_{phys}$ be still self-adjoint.)  
Generically, we expect that out of the operators corresponding to 
the set of $2D$ gauge invariant observables $(\tilde q^{a}, 
\tilde p_{a})$, we can define $D = N-M$ commuting operators, 
${\widehat O}_a$, $a=1,2,...,D$ forming a complete Dirac set.  
Assuming for simplicity these have discrete spectrum, a basis of 
physical states is labeled by their quantum numbers $n_a$, $a= 
1,2,..., D$.  A general {\it physical } state is killed by all 
the constraints ${\widehat C}_m | \psi \rangle = | 0 \rangle$, 
and can be written as
\begin{eqnarray}
|\psi\rangle & = & \sum_{n_1, ...,n_D} 
c_{n_1,...,n_D} \ | n_1, ...,n_D\rangle.  
\label{abstract states}
\end{eqnarray}
Physical evolution is described by (Heisenberg) operators 
corresponding to relational classical quantities such as 
(\ref{ec}).  In constructing these operators, ordering and 
consistency problem might, in general, be serious.


\section{Covariant statistical mechanics}

Can we use statistical mechanics methods in a covariant, 
presymplectic framework?  Energy plays an important role in 
statistical mechanics, and here there is no energy.  Statistical 
mechanics relies on the idea that systems thermalize to 
equilibrium in time.  What is thermalization in a covariant 
context, in which there is no external time variable?  To address 
these questions, our strategy will be to recall Boltzmann's logic, 
to rephrase it in the language of the presymplectic formulation 
of a conventional system, and from here, to extend it to 
presymplectic systems that do not correspond to a conventional 
system. 

Consider a Boltzmann gas in a closed box.  The gas is composed by 
a very large number $\cal N$ of identical molecules.  Begin by 
considering each molecule as free.  Let $\gamma$ be the phase 
space of a single molecule.  For instance, if we neglect 
rotational and vibrational motion, we may assume $\gamma$ to be 
six dimensional.  Since the molecule is assumed to be free, its 
motion is very simply described by a free motion in $\gamma$.  
The phase space $\Gamma$ of the entire gas has dimension $6{\cal 
N}$.  The motion of the entire gas is described by a simple 
motion in $\Gamma$ as well.  Under these assumptions, the gas 
does not thermalize, and we cannot use statistical methods.  For 
instance if we started with all the molecules bouncing up and 
down within the right half of the box, they would continue to do 
so forever, never expanding to the left-hand part of the box.  
To have thermal behavior, we need the particles to interact.  
However, taking the actual physical interaction among the 
molecules into account complicates the dynamical problem 
dramatically, and puts it far outside our theoretical capabilities.

Boltzmann's genius found a way in between, by postulating a `small,'
`thermalizing' interaction among the molecules.  The molecules bounce,
attract and repel in a non-trivial manner.  In the theoretical
description, we simply assume that each molecule is still free most of
the time, but, once in a while, it interacts with another molecule. 
We are not concerned with the details of this interaction, except for
the assumption that the interaction is maximally thermalizing, that
is, it conserves a minimal number of physical quantities.  Under this
assumption, motion in $\Gamma$ becomes ergodic and we have
thermalization.  As time goes on, the state of the gas will fill up
all allowed regions in $\Gamma$.  Of course, there are quantities that
must be conserved in any interactions, due to the homogeneity
properties of the spacetime in which the gas lives, such as momentum
and energy.  The presence of the box walls forces the total momentum
to be zero, and the only non-trivial conserved quantity is the total
energy.  Anything else is washed away by the thermalizing interaction. 
We assume that time averages are the same as ensemble averages, and
that under the action of the thermalizing interaction all microstates
of the gas become equiprobable, with the only constraint given by the
value of the total energy.  Thus macroscopic (microcanonical) states
can be labeled by a single parameter, their total energy.  As is
well known, the quantitative consequences of this very delicate
argument, considered borderline fantasy by Boltzmann's contemporaries,
are strikingly accurate in a truly impressive range of physical
contexts.

In the course of the dynamics, the motion of a single molecule 
can be followed within its phase space $\gamma$.  This motion is 
free for most of the time, but at certain times it gets suddenly 
altered: when the molecule interacts with another molecule.  
Assuming equiprobability, a simple calculation shows then that 
the time averaged distribution of the states of a single 
molecule, and thus the distribution of the molecules over the 
states, is given by $\rho \sim e^{-\beta H}$, where $H$ is the 
free Hamiltonian of the particle and $\beta$, the (inverse) 
temperature, can be computed from the total energy.

Let us now describe the same system in the presymplectic 
framework.  First, let the particles be free.  The key difference 
with the previous description is that a point in the physical 
phase space $\gamma_{ph}$ does not represent anymore the state of 
the particle at some time.  Rather, it represents a single full 
solution of the equations of motion.  (It is like a classical 
analog of a Heisenberg state, versus a Schr\"odinger state.)  
Thus, the particle motion is now described by a single, non 
moving, point in $\gamma_{ph}$, which represents a full orbit in 
$\gamma$.  Similarly, the motion of the entire gas is given by a 
single non-moving point in $\Gamma_{ph}$, corresponding to a full 
gauge orbit in $\Gamma$.  As there is no time, there is no time 
for moving around.

However, the magic, once again, happens when we turn Boltzmann's 
`small' interaction on.  The dynamics of the full system is 
still given by a single non-moving point in $\Gamma_{ph}$, or, 
equivalently, by a single orbit in $\Gamma$. However, what about the 
dynamics of a single molecule?  Since the phase space 
$\gamma_{ph}$ is defined by the {\em dynamics\/} of the system, 
and not just its kinematic as in the conventional case, it seems 
that the motion of a single molecule cannot be described in 
$\gamma_{ph}$ at all in the interacting situation, because 
$\gamma_{ph}$ is the space of the {\em free\/} motions of the 
molecule.  It seems to be a core difficulty.  However, there must 
be a way out, since, after all, we are describing the same 
physics as before.  Indeed, the way out is provided precisely by 
the assumptions about Boltzmann's thermalizing interaction.  
Observe that the orbits in $\Gamma$ do correspond to free motions 
of the single particles, interrupted by interactions.  Each such 
orbit gives, for every particle, a collection of free motions, 
namely a collection of points in $\gamma_{ph}$.  In other words, 
what the interaction does is simply make the (timeless) state in 
$\gamma_{ph}$ diffused.  A full orbit in $\Gamma_{ph}$ determines 
a distribution of points in $\gamma_{ph}$.  Under our 
assumptions, the density of these points must clearly be given by 
$\rho \sim e^{-\beta H}$!

What conclusion can we draw from this exercise of re-expressing
Boltzmann's ideas in a timeless language?  The first conclusion is
that we can still think in terms of the Boltzmann's distribution on
the states of the subsystem, even in a timeless context.  It is true
that nothing moves in the physical phase space of a fully
parameterized system, and so it seems that nothing can ever
thermalize.  But the effect of the interaction between the subsystems
can be represented precisely by a distribution on the space of
timeless non-interacting states\footnote[8]{In some sense, we are
dealing here with a `covariant ideal gas'.  As in standard ideal
gases, there is no specific interaction in the formulae involved. 
Nevertheless, the ideal gas thermalizes thanks to the Boltzmann
thermalizing interaction, which is implicitly assumed.  If specific
interaction terms among the single components $s$ were allowed, then
it would be interesting to study what thermalization could mean in
that context, as well as its relationship with the many-body forces
required to get separability of the whole system $S$ (the cluster
decomposition property) \cite{Komar}.  This issue is not addressed
here and we leave it for future developments.}.

The second conclusion regards the energy.  Why does the energy still
play a role, when the presymplectic formalism treats the time
variable, and thus the energy, just as one among other variables?  The
answer, from the above discussion, is not that the energy has any 
special importance by itself.  Rather, it is that we have simply fed
into the formalism the information that the small interaction between
the subsystems washes away everything excepts energy.  But there is
nothing sacred about energy conservation.  Energy conservation is just
a consequence of invariance under time shift, which, in turn, is a
feature of the homogeneity of the Minkowski solution under time
shifts.  We have learned from general relativity that the Minkowski
gravitational field is just one among many possible fields.  There is
no fundamental energy conservation in nature.

On the other hand, the discussion above leads us to see precisely 
under which conditions we can still use Boltzmann's statistical 
mechanics in a covariant context.  We can, anytime we have a 
system $S$ that can be seen as composed by a large number of 
identical subsystems $s$, whose dynamics is given by a free part 
which we understand well, plus a `small interaction' that can 
thermalize the macrosystem, and conserves, say, only some global 
quantities $O_{l}$.  We can formalize our conclusions as follows.

Let $(\gamma,\omega)$ be the presymplectic space describing $s$.  
Let $S$, with presymplectic space $(\Gamma,\Omega)$ be composed 
by a large number $\cal N$ of systems $s_{(n)},\ n=1,\ldots,{\cal 
N}$, all identical to $s$, having presymplectic spaces 
$(\gamma_{(n)},\omega_{(n)})$.  By this we mean 
\begin{eqnarray}
\Gamma &=& \times_{n}\ 
\gamma_{(n)},
\label{composite1}\\
\Omega &=&  \sum_{n} \omega_{(n)} + \omega_{int}\, ,
\label{composite}
\end{eqnarray}
where $\times$ indicates the Cartesian product and  
$\omega_{int}$ gives the interaction between the subsystems.  
Next, let us assume that there are $L$ quantities $O_{l}$, 
defined on $\gamma_{ph}$ (and thus on $\gamma$) such that the 
corresponding global quantities
\begin{equation}
  {\cal O}_{l}=\sum_{n} O^{(n)}_{l} \, , \quad l =1,..., L < D \, ,
\end{equation}
are invariant along the orbits of $\Omega$, that is 
\begin{equation}
X({\cal O}_{l})=0 \, , \quad l =1,..., L < D \, ,
\end{equation}
for all vector fields $X$ in $\Gamma$ such that 
\begin{equation}
\Omega(X)=0. 
\end{equation}
We call these quantities `conserved'.  Finally, we assume that 
$\omega_{int}$ suitably thermalizes all other variables besides 
the $O_{l}$'s.  This means, precisely as above, that all 
allowable (combined) states of the $s_{(n)}$ systems are, on 
average, equally covered, in moving along a generic orbit in 
$\Gamma$.

Under these conditions, we can straightforwardly construct a covariant
statistical formalism.  A state in $\Gamma_{ph}$, determines, for each
$s_{(n)}$ a distribution $\rho$ on $\gamma_{ph}$, which gives the
distribution of `initial states' of the component system as we move
around the corresponding orbit of the interacting composite system. 
For a generic state in $\Gamma_{ph}$, this distribution can be
computed using conventional statistical techniques, in particular, by
assuming that the distribution is the one that maximizes the number of
possible microstates compatible with the given macrostate.  The
result is straightforward: the (unnormalized) probability distribution
on the phase space is
\begin{equation}
\rho = e^{-\gamma^l {O}_l}, 
\end{equation}
where  $\gamma^l$ are the intensive thermodynamical parameters 
that determine the equilibrium of the members of the 
ensemble with respect to the transfer of the quantities 
$[{\widehat O}_l]$.   

Instead of detailing the classical theory, we discuss directly the
quantum theory.  We use von Neumann's density operator formalism
\cite{Neumann}.  We ask the ensemble to satisfy a maximum entropy
principle.  In other words, we ask the quantum statistical entropy $S$
per constituent member of the ensemble given by
\begin{eqnarray}
S = - k\ \mbox{Tr}\,\, \left ( {\widehat\rho} 
\ln{\widehat \rho} \right )\, ,
\end{eqnarray}
to be a maximum under the constraints
\begin{eqnarray}
\mbox{Tr}\,\,( {\widehat \rho}\,{\widehat O}_l )  & = &
{\bar O}_l ,\nonumber\\
\mbox{Tr}\,\,{\widehat \rho} & = & 1 \, , \label{VALUES}
\end{eqnarray}
where ${\widehat O}_l$ are the quantum operator corresponding to 
the conserved quantities ${O}_l$ and ${\bar O}_l$ are fixed 
average values.  ${\widehat \rho}$ is the density operator.  The 
density operator ${\widehat \rho}$ that fulfills these requirements is
\begin{eqnarray}
{\widehat \rho} & = &  {\cal Z}^{-1}\ e^{-\gamma^l {\widehat 
O}_l}, 
\end{eqnarray}
with 
\begin{eqnarray}
{\cal Z} = \mbox{Tr}\,\, e^{-\gamma^l {\widehat O}_l}\, , 
\end{eqnarray}
the partition function.  The thermodynamical parameters $\gamma^l$ 
can be obtained from the conditions (\ref{VALUES}) provided that 
the matrix $\frac{\partial [{\widehat O}_i ] }{\partial 
\gamma^j}$ have non-vanishing determinant.  Clearly, they are the 
parameters that measure the equilibrium of the members of the 
ensemble with respect to the transfer of the quantities 
$[{\widehat O}_l]$.  

In the conventional case of a non-covariant system formulated in
covariant terms, only one non-trivial quantity is conserved, the
energy, and we obtain the standard results.  In particular, we can
consider a gravitational system with an asymptotically flat
gravitational field; in this case the Hamiltonian is given by suitable
boundary terms and the observables at infinity evolve in the Lorentz
time of the asymptotic metric.  The system is Lorentz invariant for the
asymptotic Lorentz transformations, and therefore, in particular,
invariant for time translations. Therefore all interactions conserve 
the asymptotic Lorentz energy and the theory considered here reduces 
to the standard results.  A specific example of this is given by all 
the literature on black hole thermodynamics an statistical mechanics, 
in which, in general, the gravitational field is assumed to be 
asymptotically flat.

\section{An example}

As a simple example, we take as component system $s$ a model with 
two non-commuting Hamiltonian constraints and one physical degree of freedom 
which was studied in \cite{Mon99}.  This model mimics the 
constraint structure of general relativity.  We refer to 
\cite{Mon99} for all details. 

The model we consider is defined by the action 
\begin{eqnarray}
S[{\vec u},{\vec v}, N, M , \lambda ] = \frac12 
{\displaystyle \int} dt \left[\, N\, ({\cal D} \vec u^2 + \vec v^2)
+ M\, ({\cal D} \vec v^2 + \vec u^2)\, 
\right], 
\end{eqnarray}
where
\begin{equation}
{\cal D} {\vec u}  =  \frac{1}{N} (\dot {\vec u} - 
\lambda {\vec u}), \ \ \ \ \ 
{\cal D} {\vec v}  =  \frac{1}{M} (\dot {\vec v} + 
\lambda {\vec v}); 
\end{equation}
the two Lagrangian dynamical variables ${\vec u}=(u^1,u^2)$ and 
${\vec v}=(v^1,v^2)$ are two-dimensional real vectors; $N$, $M$ 
and $\lambda$ are Lagrange multipliers.  The squares are taken in 
$R^{2}$: $\vec u^2 = \vec u\cdot \vec u = 
(u^{1})^{2}+(u^{2})^{2}$.
The action can be put in the form (\ref{act}),
\begin{eqnarray}
S [ \vec u, \vec v, \vec p, \vec \pi, \lambda^m ] =  
\int d \tau \,\, \left \{
\dot{\vec u}\cdot \vec p+
\dot{\vec v} \cdot \vec \pi 
- \lambda^m C_m (q^i,p_i) \right \} \, . 
\end{eqnarray} 
The canonical variables $(\vec u, \vec v, \vec p, \vec \pi)$ 
define the eight dimensional extended phase space $\gamma_{ex}$, 
with symplectic form $\omega_{ex} = d\vec p\wedge d\vec u+ d\vec 
\pi\wedge d\vec v$, also $\lambda^1 = N$, $\lambda^2 =M$, and 
$\lambda^3 =\lambda$.  The constraints 
\begin{eqnarray}
C_1 & = & \frac12 (\vec p^2 -\vec v^2)\, ,\nonumber\\
C_2 & = & \frac12 (\vec{\pi}^2 -\vec u^2)\, ,\nonumber\\
C_3 & = & {\vec u}\cdot {\vec p} - 
{\vec v} \cdot {\vec \pi} \, ,  \label{Hamiltonian} 
\end{eqnarray}
are first class and define a five dimensional constraint 
surface $\gamma$ in  $\gamma_{ex}$, and 
$\omega=\omega_{ex}|_{\gamma}$. 

A complete set of gauge-invariant quantities is given by the two 
continuous quantities $J\in R^{+}, \phi\in S_{1}$ and two 
discrete quantity $\epsilon,\epsilon'=\pm 1$, defined by
\begin{eqnarray}
	\epsilon & = & {u^{1}p^{2}-p^{1}u^{2}\over|u^{1}p^{2}-p^{1}u^{2}|} 
	\, , \nonumber \\
	\epsilon' & = 
	&{\pi^{1}v^{2}-v^{1}\pi^{2}\over|\pi^{1}v^{2}-v^{1}\pi^{2}|} \, , 
	\nonumber \\
	J & = & |u^{1}p^{2}-p^{1}u^{2}| \, , 
	\nonumber  \\
	\phi & = & \arctan{u^{1}v^{2}-p^{1}\pi^{2}\over
	u^{1}v^{1}-p^{1}\pi^{1}}\, , 
	\label{jpe}
\end{eqnarray}
These can be taken as coordinates of the physical gauge-invariant 
phase space.  The quantity $J$ resembles an angular momentum, 
and thus it is called as such.

Let us now consider a large number of systems of this kind which are 
interacting weakly.  The composite system dynamics is given by 
the presymplectic system (\ref{composite}).  Let us assume, as an
example, that $\omega_{int}$ is a sum of binary interactions in 
which the sum of the two angular momenta, while all other 
quantities are thermalized.  This defines the statistical 
mechanics of the composite system. 

In the quantum theory, we take a complete Dirac set of commuting 
operators ${\widehat J}$, ${\widehat \epsilon}$ and ${\widehat 
{\epsilon'}}$.  Their spectrum, worked out in \cite{Mon99}, is
\begin{eqnarray}
{\widehat J} |m,\epsilon , \epsilon' \rangle_N & = & J_m 
|m, \epsilon , \epsilon' \rangle_N = m \hbar  
|m, \epsilon , \epsilon' \rangle_N \, ,\nonumber\\
{\widehat \epsilon} |m,\epsilon , \epsilon' \rangle_N & = & 
{\epsilon} |m, \epsilon , \epsilon' \rangle_N \, ,\nonumber\\
{\widehat {\epsilon'}} |m,\epsilon , \epsilon' \rangle_N & = & 
{\epsilon'} |m, \epsilon , \epsilon' \rangle_N \, .
\end{eqnarray}  
$m$ is a positive integer, and $\epsilon$ and $\epsilon'$ take 
values 1 and -1.  The states $|m,\epsilon , \epsilon' \rangle_N$ 
form a normalized basis in the physical Hilbert space of the 
theory.
\begin{eqnarray}
_N\langle m,\epsilon , {\epsilon}'  |{\tilde m},{\tilde \epsilon} , 
{\tilde \epsilon}' \rangle_N = \delta_{m , {\tilde m'}} 
\delta_{\epsilon , {\tilde {\epsilon}}} 
\delta_{{\epsilon}' , {\tilde \epsilon}'}\, .
\end{eqnarray}
The quantity $J$ is represented by the operator ${\widehat J}$, 
which is a kind of angular momentum \cite{Mon99} as we have 
mentioned.  If we assume that the thermalizing interaction 
conserves the total value of $J$, we have immediately the density 
operator
\begin{eqnarray}
{\widehat \rho} & = & \frac{e^{-\gamma {\widehat J}}}{\cal Z}\, .
\end{eqnarray}
Using $\mbox{Tr}\, {\widehat \rho}=1$, we get the partition
function
\begin{eqnarray}
{\cal Z} = \sum^{\infty}_{m=1} \omega_m e^{-\gamma J_m}= 
\frac{4}{ e^{\gamma \hbar} -1 }\, , 
\end{eqnarray}
with $\omega_m=4$ because there are 4 states for a given $m$, and 
also $e^{-\gamma \hbar}<1$ has been used (i.e., positive 
`temperature' $\gamma$ has been assumed).  The angular momentum 
per constituent is given by 
\begin{eqnarray}
L := [{\widehat J} ] & = & \frac{\mbox{Tr}\,\, ( e^{-\gamma 
{\widehat J}} \, {\widehat J} ) } {\cal Z} = 
-\frac{\partial}{\partial \gamma }\ln{\cal Z} = \frac{e^{\gamma 
\hbar} \hbar }{(e^{\gamma \hbar} -1)} \, ,
\end{eqnarray}
and the entropy per constituent 
\begin{eqnarray}
S & = & k \sigma = - k \mbox{Tr}\,\, \left ( {\widehat\rho} 
\ln{\widehat \rho} \right ) = k \ln{\cal Z} + k \gamma L\, .
\end{eqnarray}
The parameter $\gamma$ characterizes the equilibrium state of the 
system with the reservoir and plays here the role of a 
temperature.  If we had an empirical thermodynamics of this 
system, we could identify this parameter with an empirically 
determined thermodynamical quantity.



\section{Gas of free relativistic particles}

Before concluding, we discuss a simple case, in which some of the
ideas presented above can play a role: the case of a gas of free 
relativistic particles.  This is really an oversimplified
situation, which can be treated with simpler tools; but the
illustration of this case may be instructive, and can be seen as
a check that the theory described here is in agreement with other
methods in the cases in which other methods can be applied. For 
simplicity, we remain here in the classical context. 

Consider thus a gas of relativistic particles.  Can we associate
a temperature to this gas?  What is the statistical state
describing these particles?  The Hamiltonian description of a
single relativistic particle can be formulated in a manifestly
Lorentz covariant fashion as follows.  The phase space is
coordinatized by the coordinates $x^{\mu}$ and their conjugate
momenta $p_{\mu}$ --that is, the symplectic form is
$\omega_{ext}=dx^{\mu}\wedge dp_{\mu}$-- and the dynamics is
given by the constraint $C=p^{2}-m^{2}$.  The seven dimensional
constraint surface $\gamma$ defined by $C=0$ with its induced
restriction $\omega$ of $\omega_{ext}$ form the presymplectic
space $(\gamma,\omega)$ describing the dynamics of the particle.

Now, one may say that for this system we know that the time is
$t=x^{0}$, and the energy is $E=p_{0}$.  Therefore we can apply
standard statistical mechanics with no difficulties.  However, there
are two distinct problems.  The first is that the entire dynamics of
the system is contained in the geometry of $\gamma$, which has no a
priori specification of which variable is time and which variable is
energy.  Thus, can we do thermodynamics just on the basis of the
actual dynamical laws, without specifying which one is the time
parameter? However, there is also a second problem, much more concrete. 
Suppose we say that $x^{0}$ is the time variable, and $p_{0}$ is the
energy.  You, on the other hand, use a different Lorentz reference
frame, and therefore for you the time is $x'{}^{0} = \Lambda^{0}_{\mu}
x^\mu$ and the energy is $p'_{0} = \Lambda^{\mu}_{0}p_{\mu}$, where
$\Lambda$ is a Lorentz transformation.  If I write a Boltzmann 
statistical state using my definition of energy, and you in yours, do 
we define the same statistical state? It is easy to see that the 
answer is no, because 
\begin{equation}
    \rho(x,p) = e^{-\beta p_{0}}\ \ \ne \ \ \rho'(x,p) =
    e^{-\beta' p'_{0}}.
    \label{eq:distripart2}
\end{equation}
whatever is $\beta'$. 
So, which one is the correct equilibrium state? 

Let us address both problems in terms of the general theory developed
above.  The key point is that if the gas is formed by particles that
are really free, they will never thermalize.  Some thermalizing
interaction is needed in order to reach an equilibrium state.  Thus,
we need some additional physical input (this is the key point).  On
physical grounds, we may for instance observe that our gas of
relativistic particles thermalizes by means of relativistic elastic
scattering. Therefore, the dynamics of a single particle is not really
free: the presymplectic space describing the dynamics of the system is
the cartesian product of the spaces of the particles, the total
presymplectic form is the sum of the individual presymplectic forms
plus the interaction term, as in Eqs.(\ref{composite1} and 
\ref{composite}).  What are the conserved quantities $O_{l}$, in the
sense of Section III? Namely, what are the quantities that are
exchanged, but whose total value is conserved, in such an interaction? 
Clearly, they are the momenta $p_{\mu}$.  Therefore, according to the
general theory of Section III, the (unnormalized) probability
distribution of the states of the single particle is
\begin{equation}
    \rho(x,p)=e^{-\gamma_{\mu}p^\mu},
    \label{eq:rhogp}
\end{equation}
where the quantities $\gamma_{\mu}$ are the intensive parameters 
describing the system. Straightforward application of standard 
statistical techniques tells us then that the average 4-momentum 
is proportional to  $\gamma^{\mu}$: 
\begin{equation}
    \bar p^\mu  =\frac{\int_{C} dxdp\  p^\mu \ 
    e^{-\gamma_{\mu}p^\mu}}{\int_{C} dxdp\  e^{-\gamma_{\mu}p^\mu}}
    = -\frac{d}{d\gamma_{\mu}}\ln \int_{C}dp\ e^{-\gamma_{\mu}p^\mu}
    = \frac{\gamma^{\mu}}{|\gamma|^{2}}. 
    \label{eq:gamma}
\end{equation}
Therefore, $\gamma^{\mu}$ must be timelike, and therefore there is a 
preferred Lorentz frame in which  $\vec\gamma=0$ and  
$\gamma_{0}=\beta$. This is the frame in which the center of mass of 
the cloud is at rest.  In this frame, the 
(unnormalized) probability distribution of the states
of the single particle is 
\begin{equation}
    \rho(x,p)=e^{-\beta p_0}. 
    \label{eq:distripart1}
\end{equation}
We can learn various lessons from this.  First, there is no Lorentz
invariant thermal state: a thermal state is in equilibrium in a {\em
preferred\/} Lorentz frame of reference, and therefore breaks Lorentz
invariance.  Second, we can say that the form of the statistical state
is physically determined by the fact that the thermalizing interaction
conserves $p^\mu$, and not by the fact that preferred phase space
coordinates play an a priori role of time and energy.

The argument above can indeed be sharpened by a more detailed analysis
of the physics of the system. Note that if the only interaction is
elastic scattering among the particles, then the gas will diffuse and
fail to reach equilibrium.  Thus, we need something that keeps the gas
contained, in order to have a meaningful thermodynamics.  One
possibility to keep the gas contained is to put it in a box.  The
position of the box will then break Lorentz invariance and pick the
preferred Lorentz frame in which the box does not move. 
Alternatively, we may think that the particles are gravitationally
bound. To have Lorentz invariance, we need a field theory for the
gravitational field.  (We can disregard here the difficulties of
having point particles, or rigid particles, in general relativity,
which play no role in this context).  In this case, we can approximate
the dynamics of a single particle as the dynamic of a particle in the
mean gravitational field of the others.  At equilibrium, this
gravitational field will be stationary in a a preferred Lorentz frame,
the one of the center of mass of the cloud.  Now, in both cases a
single particle is not longer free, but rather is subjected to an
interaction which is not longer Lorentz invariant.  In fact, in both
cases the interaction preserves energy but not momentum (in the second
case, energy is kinetic plus potential.)  Therefore in both cases
there is a preferred $p_{0}$ which is conserved in the course of the
thermalizing interaction.  It is clearly {\em this\/} energy the one
that enters Eq.(\ref{eq:distripart1}), because this is the energy
which is totally conserved and freely exchanged in the system, and
thus which becomes equipartioned.  In other words, this is the quantity
that is conserved in the thermalizing interaction, and that becomes
the extensive thermodynamical parameter governing the system.

The general lesson should be clear at this point.  Even if we do not
have an a priori recognition of which function on the extended phase
space represents time (or energy), we can nevertheless run the
statistical mechanics formalism on the basis of the quantities that
are preserved in the thermalizing interaction.


\section{Conclusions and perspectives}

We have argued that quantum statistical mechanical techniques can 
be applied to a macroscopic generally covariant 
system composed of a large number of generally covariant 
subsystems.  This can be done without arbitrarily selecting a 
variable as the time variable, and in spite of the absence of a 
notion of energy.

We recall that in the literature there are two main schools of thought
in relation to the `problem of time' in generally relativistic
theories.  One tries to single out the `correct'\, time variable among
the variables of the covariant theory.  The choice determines a
preferred Hamiltonian (energy), and thus an unambiguous concept of
temperature.  The opposite point of view, which we consider more
fruitful and we have developed here, takes general covariance more
seriously, and keeps all variables on the same footing.  From this
point of view, temperature plays no fundamental role in the
statistical analysis.  It may not be defined, or, if it is defined at
all, temperature is just one of the intensive macroscopic parameters
characterizing the equilibrium configuration of the system.

Our basic idea is that if the macroscopic system can be viewed as being 
formed from weakly interacting systems, then a full solution of the 
equation of motion of the macroscopic system determines a 
distribution of solutions of the equations of motion of the 
components.  The properties of the interaction determine which 
global quantities are conserved and thus the extensive 
macroscopic parameters.  In turn, these determine intensive 
thermodynamical parameters that describe the macrostate.  The 
other microscopic degrees of freedom are thermalized away by the 
interaction.  The fact that a preferred single notion of 
temperature does not necessarily arise is not surprisingly, given 
the weak and always contingent role that energy plays in general 
relativistic theories.  The statistical mechanics of generally 
covariant theories does not depend on the notion of energy.  
Rather, ensembles are determined by the properties of the 
thermalizing interaction.

What is a thermometer in this context?  In the usual context a
thermometer is a physical system which has the property of having a
macroscopic variable $h$ (the height of the mercury column) directly
coupled to the average energy.  By looking at $h$ we measure directly
the average energy and therefore the temperature.  If local energy is
not conserved then a conventional thermometer will keep measuring its
own average energy, but this will give little information on the
system, because individual subsystems will not thermalize to the same
reading of the thermometer.  On the other hand, if other intensive
parameters $\gamma^{l}$ are conjugate to other conserved quantities
$O_{l}$, then in principle specific ``thermometers" measuring the
average value of $O_{l}$ may exist.  These should play the same role
as conventional thermometers.

Our approach has been very abstract, and applications in 
realistic general relativistic contexts may not be trivial.  A 
first naive idea is to obtain a system composed of subsystems by 
partitioning space into small patches, each with its own 
gravitational degrees of freedom.  This procedure, however, might 
interfere badly with general covariance.  The example considered 
in the text suggests to look at the strong coupling limit of 
general relativity, which is precisely given by a collection of 
finite dimensional covariant systems.  In this context, we recall 
that near singularities --such as the cosmological one-- notions 
of temperature and entropy are usually very badly defined.  
Alternatively, one could think of somehow Fourier expanding the 
gravitational field, and partitioning it in momentum space, 
following fluid techniques.  Perhaps a realistic context in which 
a covariant statistical mechanics may find application is where 
matter and strong gravitational fields are both present.  The 
presence of matter could lead to a natural physical way of 
partitioning the degrees of freedom.  In general, any context in 
which thermal energy can be lost substantially in the 
gravitational field would, in principle, require a covariant 
thermodynamics.

On the purely theoretical side, a natural open issue is the 
quantum statistical mechanics of constrained systems with 
symmetries in the global quantum states.  That is, the covariant 
version of Fermi-Dirac, and Bose-Einstein statistics.  

The relation between coherent states in standard quantum 
mechanics and thermodynamics \cite{Lieb1973} suggests that 
coherent state quantization of constrained systems might bring a 
better understanding of the thermodynamics of generally covariant 
systems \cite{Klauder97}, and also shed light on the tantalizing 
issue of the classical limit states in quantum gravity.  In 
particular, consider the spacetime described by a (generic) 
statistical mixture of gravitational states $|\psi\rangle$.  In 
loop quantum gravity, $| \psi\rangle$ are superpositions of 
$s$-knots, or abstract spin networks.  In the basis in which 
${\widehat \rho}$ is diagonal, we have
\begin{eqnarray}
{\widehat \rho} = \sum_{\psi} P(\psi) \mid \psi \rangle 
\langle \psi \mid
\end{eqnarray}
and we can compute, for instance, the density matrix yielding 
average macroscopic values of the geometry.  In particular, we 
can use area and volume operators associated with a compact 
region of space and require something like 
\begin{eqnarray}
{\bar A} = \mbox{Tr}\,\, 
{\widehat \rho} {\widehat A}\, , \nonumber\\
{\bar V} = \mbox{Tr}\,\, 
{\widehat \rho} {\widehat V}\, .
\end{eqnarray}
This may determine a statistical state of the kind 
\begin{eqnarray}
{\widehat \rho} = \frac{e^{-\alpha {\widehat A} -\beta {\widehat 
V}}} {\cal Z} \, , 
\end{eqnarray}
which might describe the physical state of spacetime better than 
a somewhat arbitrary pure state. 

In closing, let us emphasize again that we have not discussed here the
statistical mechanics of matter interacting with a fixed gravitational
field.  Rather, we have considered the full quantum statistical
mechanics of spacetime itself.  Similarly, we have not discussed the
statistical mechanics of black holes, which focuses on (the
perturbations of the gravitational field around) certain preferred
black hole configuration, and is generally in the asymptotically flat
context, in which boundary quantities determine time and conserved
energy.  Finally, we think that the proposal of the statistical and
algebraic origin of the time flow \cite{Rovelli:1993,Connes:1994}
might be reconsidered at the light of the physical ideas discussed
here.  The magic secret coffer of the relations between gravity,
quantum and thermodynamics is still far from being fully open.


\section*{Acknowledgments} 
Warm thanks to Ted Newman for many discussions and many helpful
criticisms.  Thanks to an anonymous referee for useful and detailed
criticisms to the first version of this work.  MM thanks financial
support provided by the {\it Sistema Nacional de Investigadores} (SNI)
of the Secretar\'{\i}a de Educaci\'on P\'ublica (SEP) of Mexico.  The
summer stay of MM at the Department of Physics and Astronomy of the
University of Pittsburgh, where this paper was finished, is supported
by the {\it Mexican Academy of Sciences} and {\it The United
States-Mexico Foundation for Science}.  Also MM thanks all the members
of the Department of Physics and Astronomy of the University of
Pittsburgh for their warm hospitality.  This work was partially
supported by NSF grant PHY-9900791.


\end{document}